\documentclass[usenatbib,conference]{basi}
\usepackage[T1]{fontenc}
\usepackage[british]{babel}
\usepackage[varg]{txfonts}
\usepackage{rotating}
\usepackage{dcolumn}
\begin{document}
\title[Making MILES better]{Making MILES better for stellar population modelling}
%\thanks{By Andr\'e Milone, email: \texttt{andre.milone@inpe.br}}}
%
\author[A.~de~C.~Milone~et~al.]
{A.~de~C.~Milone$^1$\thanks{email: \texttt{andre.milone@inpe.br}},\\
A.~Sansom$^{2}$,
A.~Vazdekis$^{3,4}$,
P.~S\'anchez-Bl\'azquez$^{5}$,
C.~Allende~Prieto$^{3,4}$,
J.~Falc\'on-Barroso$^{3,4}$
and R.~da~Silva$^{6}$\\
%$^1$Divis\~ao de Astrof{\'\i}sica, Instituto Nacional de Pesquisas Espaciais, Av. dos Astronautas 1758,\\
%12227-010, S\~ao Jos\'e dos Campos, Brazil\\
$^1$Instituto Nacional de Pesquisas Espaciais, Av. dos Astronautas 1758, 12227-010, S. J. Campos, Brazil\\
$^2$Jeremiah Horrocks Institute, University of Central Lancashire, Preston, PR1 2HE, UK\\
$^3$Instituto de Astrof{\'\i}sica de Canarias, V{\'\i}a L\'actea s/n, E-38205, La Laguna, Tenerife, Spain\\
$^4$Depto. de Astrof{\'\i}sica, Universidad de La Laguna, E-38206, La Laguna, Tenerife, Spain\\
$^5$Depto. de F{\'\i}sica Te\'orica, Universidad Aut\'onoma de Madrid, Cantoblanco 28409, Madrid, Spain\\
$^6$Osservatorio Astronomico di Roma, Via Frascati 33, 00040, Monte Porzio Catone, Italy}

\pubyear{2014}
\volume{10}
\pagerange{\pageref{firstpage}--\pageref{lastpage}}
%\status{submitted}

\date{Received --- ; accepted ---}

\maketitle
%------------------------------------------------------------------------------%
% abstract and keywords                                                        %
%------------------------------------------------------------------------------%
\label{firstpage}

\begin{abstract}
In order to build more realistic single stellar population (SSP) models with variable $\alpha$-enhancement,
we have recently determined [Mg/Fe] in a uniform scale with a precision of about 0.1 dex
for 752 stars in the MILES empirical library.
The [$\alpha$/Fe] abundance ratio is commonly used as a good temporal scale indicator of star formation,
taking Mg as a template for $\alpha$ elements.
Calcium is another element whose abundance is currently being investigated for the MILES stars.
The MILES library is also being expanded by around 20\% by including stars with known
{\it T}$_{\rm eff}$, log~$g$, [Fe/H] and [Mg/Fe].
The transformation of their photospheric parameters to the MILES system
has been carried out, but the calibration of their [Mg/Fe] is still in progress.
In parallel, C, N and O abundances are also being compiled from literature for the library stars
because they play an important role in the photospheric opacity, particularly influencing the blue spectral region. 
The Galactic kinematic classification of MILES stars with compiled [Mg/Fe] has been just computed
such that this information can be considered in the SSP modelling.
Comparisons of theoretical stellar predictions of the Lick line-strength indices against the MILES data
have revealed the good behaviour of Fe-sensitive indices predictions,
while highlighting areas for improvement in some models for the higher order H-Balmer features.
%\\[6pt]
\end{abstract}


% Section 1 ------------------------------------------------------------------------------
% main text of the paper, using \section, \subsection, \subsubsection ...
%------------------------------------------------------------------------------
\section{Introduction}\label{s:intro}

The main limitation of current empirical stellar spectral libraries is the use of stars
whose individual elemental abundances are not adequately considered.
Moreover, the stars in these libraries are basically selected from the solar neighbourhood,
in which the chemical evolution of Milk Way is imprinted. 
That limitation is also propagated to the SSP isochrone-based models that have been built on these libraries.
Typically, only iron is taken as a metallicity tracer, but the spectral energy distributions of stars
and stellar systems considerably depend on abundance ratios of other metals relative to iron
(e.g. CNO group and alpha elements).
For instance, [$\alpha$/Fe] is known as a good temporal scale indicator for the star formation
since the alpha and iron-group elements have distinct nucleosynthetic origins with different time scales,
respectively type II and type Ia supernovae.
The higher this ratio is, the shorter the time scale for stellar formation.
Magnesium is usually assumed as a template for $\alpha$ elements.

Therefore we have been taking steps to improve one of the most widely used empirical stellar spectral libraries
- MILES \citep{2006MNRAS.371..703S} \& \citep{2007MNRAS.374..664C},
and to use it to test theoretical models.
Our ultimate aim is to utilise a combination of well characterised stars and theoretical model spectra
to predict SSP spectra for a wide range of abundance patterns.
Furthermore, building a new generation of semi-empirical SSP models with variable [$\alpha$/Fe]
will open new possibilities to minimise the effect of extrapolations of [$\alpha$/Fe] in the models parametric space
as well as to break the age-metallicity degeneracy over the model line-strength predictions,
improving the recovery of star formation history in galaxies.

% Section 2 --------------------------------------------------------------------%
\section{[E/Fe] abundance ratios in MILES stars: Mg, Ca and CNO}

[Mg/Fe] was determined with a systematic uncertainty of about 0.1 dex
for 752 stars in the MILES empirical library \citep{2011MNRAS.414.1227M}.
This abundance ratio was compiled from high-resolution ({\sl HR}) spectroscopic analyses
for 315 stars (263 dwarfs and 52 giants) with a systematic error of 0.09 dex on average.
437 stars (150 dwarfs and 287 giants) had the ratio measured at mid spectral resolution
with an average systematic error of 0.12 dex
by carefully analysing two Mg~I atomic features ($\lambda$5183{\AA} and $\lambda$5528{\AA})
in the own MILES spectra through two almost independent methods
(pseudo equivalent width, p-EW, and line profile fit, LPF).
The mid-resolution ({\sl mr}) analysis was based on spectral synthesis solved
with the LTE radiative transfer code
MOOG\footnote{Sneden C., 2002, The MOOG code, http://verdi.as.utexas.edu/moog.html},
adopting the standard composition group of MARCS model atmospheres
that follow the Galactic global trend of [$\alpha$/Fe] as a function of [Fe/H] \citep{2008A&A...486..951G}
together with VALD atomic \citep{2000BaltA...9..590K}
and \citet{1995ASPC...81..583K} molecular absorption line lists.
The assumed solar abundance pattern was that adopted by the MARCS 2008 models
\citep{2007SSRv..130..105G}.
All the [Mg/Fe] values were calibrated to a single uniform scale,
by using an extensive control sample with results from {\sl HR} studies.
Rather than be uniform over all photospheric parameter scales,
the {\sl mr} measurements actually complement the {\sl HR} data.
Depending on the parametric region, one Mg feature works better than the other.
In general Mg~I~$\lambda$5183 gives reliable results for dwarfs/giants with {\it T}$_{\rm eff}$ between 4000 and 8000 K,
while Mg~I~$\lambda$5528 basically works on dwarfs with 3500 up to 8000 K
and giants for 3600 $\leq$ {\it T}$_{\rm eff}$ $\leq$ 5500 K.
See more details in \citet{2011MNRAS.414.1227M}.

Calcium is another $\alpha$ element whose abundance is currently being investigated for the library stars.
So far we have collected [Ca/Fe] from a few {\sl HR} works for dozen stars
as well as we have just started analysing at mid-resolution a single Ca~I atomic feature ($\lambda$5513{\AA})
to check the precision of the resulting abundance ratios through the p-EW and LPF methods
as a function of the local spectrum S/N ratio
(an example of spectral synthesis is shown in Figure 1).
Other prominent Ca features will be further analysed in the MILES spectra
and an extensive search for {\sl HR} data will be performed as well.

% Figure 1
\begin{figure}
\centerline{\includegraphics[width=7.00cm, angle=-90]{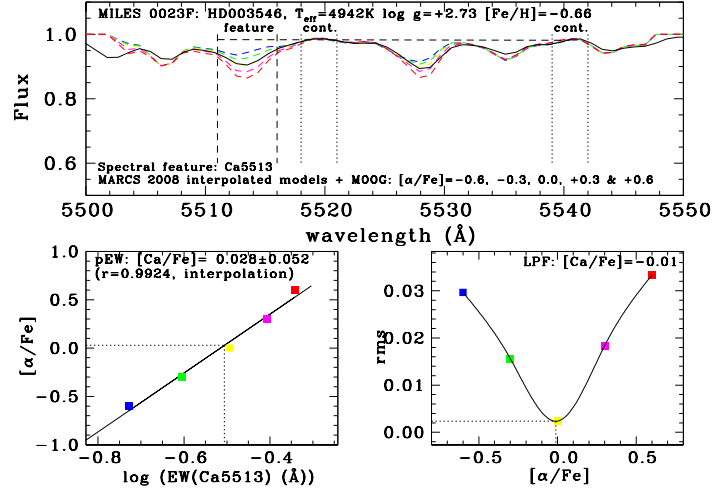}}
\caption{Example of spectral synthesis of the Ca~I~$\lambda$5513 feature for a MILES star.}
\end{figure}

C, N and O abundances are also being compiled from published {\sl HR} works for the MILES stars
because these elements play an important role in specific ranges of optical stellar spectra,
including strong features blue-ward of about 4400{\AA}.
We have also been investigating which {\sl HR} work(s) will be chosen as reference(s)
to define uniform scales of the [E/Fe] abundance ratios for Ca, C, N and O.
The ideal solution would be to find out a single {\sl HR} study
that provide abundances for all four element for a common sample within MILES
covering a range in {\it T}$_{\rm eff}$, log~$g$ and [Fe/H] as wide as possible,
such that the work's photospheric parameters should have acceptable agreements
with those compiled values in MILES under a given statistical criterion.

% Section 3 -----------------------------------------------------------------------%
\section{Comparisons against stellar spectral models: Lick indices}

With the iron and magnesium abundance characterisations it was possible to use stars in the MILES library
to test some of the theoretical predictions of Lick System line-strength indices available in the literature,
for their spectral responses to changes in element abundances.
These are known as theoretical response functions and consist of tables of changes in Lick spectral indices,
due to doubling of individual elemental abundances.
The baseline models typically have solar photospheric abundance patterns.
Such tables of response function have been used by many people to estimate Lick indices in stellar populations,
allowing for non-solar abundance patterns.
\citet{2013MNRAS.435..952S}
describe our tests of published response functions
and discusses their applications and limitations.
That paper was the first attempt to test the theoretical predictions against results
from a library of observed stellar spectra. 

Three star models in particular were tested, for a cool giant, turn-off and cool dwarf star.
We tested 23 Lick indices using predicted changes in 9 elements and in overall metallicity.
This work showed good agreement for some Lick indices, but poor agreement
for the H{$\gamma$} and H{$\delta$} indices,
which were predicted to cover a broader range of values than observed, in cool stars;
see figures 1 \& 2 of \citet{2013MNRAS.435..952S}.
Such spectral indices are useful for age dating unresolved stellar populations,
so in future work we will test these indices,
as new databases of theoretical stellar spectral models become available.
Better self consistency between element abundance changes and atmospheric opacities
may be needed to improve the theoretical model predictions for variable abundance patterns.
It is hoped that such tests will prove valuable both to those who make the models
and to those who apply them in order to gain understanding of the evolutionary history of the stars in galaxies.

Continued efforts to accurately characterise the MILES stars will add to their usefulness
for testing models of specific stars,
in different parts of the Hertzsprung-Russell diagram.
To this end we are compiling abundances available from high resolution spectra,
for key elements such as C, N, O
and for other alpha elements such as Ca (see Figures 2, 3 and 4).

% Figure 2
\begin{figure}
\centerline{\includegraphics[width=5.88cm, angle=-90]{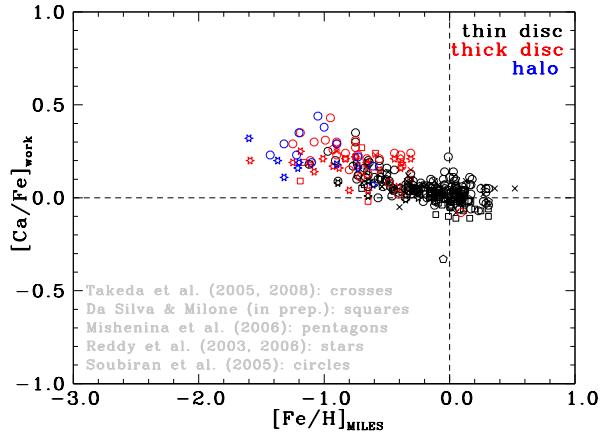}}
\caption{[Ca/Fe]$_{\rm work}$ vs. [Fe/H]$_{\rm MILES}$
(Galactic kinematic groups distinguished with different colours).}
\end{figure}

% Section 4 ------------------------------------------------------------------------------%
\section{Metallicity effects on blue spectral region}\label{s:maths}

Comparisons of MILES spectra between pairs of similar stars (in {\it T}$_{\rm eff}$ and log~$g$)
for three different evolutionary stages
(cool main-sequence dwarf at {\it T}$_{\rm eff}$ = 5100 K, main-sequence turn-off for 4 Gyr SSP
and red normal giant at {\it T}$_{\rm eff}$ = 4000 K)
have highlighted the strong influence of iron abundance blue-ward of 4400{\AA}
\citep{2013MNRAS.435..952S}.
Also, this region of the spectrum contains strong molecular features
due directly or indirectly to C, N and/or O.
These include the CN and CH bands (e.g. G band at $\lambda$4300{\AA})
that can be quantified through line-strength indices such as CNO3862 and CNO4175
\citep{2005ApJ...627..754S}.
The flux excess in these features for $\alpha$-enhanced red giants
in comparison with less $\alpha$-enriched ones at a fixed overall metallicity [Z/H] occurs mainly
because of a decrease in [Fe/H] instead of an increase in [$\alpha$/Fe].
The impact of these abundance variations on the blue spectral region
decreases with increasing temperature as expected.
These results are not influenced by the uncertainties in photospheric parameters.
Overall [Z/H] is not always the most appropriate way to rank stars,
because abundances of individual elements such as Fe, C and N have important
effects on blue emergent stellar spectra, particularly important in cool stars.

% Section 5 ------------------------------------------------------------------------------%
\section{Kinematics properties of MILES stars}\label{s:maths}

We have classified the MILES stars according to their main Galactic population membership
by computing the probability of a given star belonging to the thin disc, thick disc, halo,
or to a transition population.
First, using equations from \citet{1987AJ.....93..864J} applied to parallaxes, proper motions, and radial velocities,
we have computed the space-velocity components $U$, $V$, and $W$
with respect to the Local Standard of Rest (LSR).
For the Sun we adopted $U_{\rm LSR}$ = 10, $V_{\rm LSR}$ = 5.3,
and $W_{\rm LSR}$ = 7.2~km\,s$^{-1}$ \citep{1998MNRAS.298..387D}.
Then, applying the equations of \citet{2006MNRAS.367.1329R}
to the derived space-velocity components,
we have computed the probability that a star belongs either to one of the three populations
(for a probability $P$ $\geq$ 70\%) or to a transition population.
The resulting classification is accordingly shown in the plots [E/Fe] vs. [Fe/H] of Figures 2, 3 and 4.

% Figure 3
\begin{figure}
\centerline{\includegraphics[width=5.90cm, angle=-90]{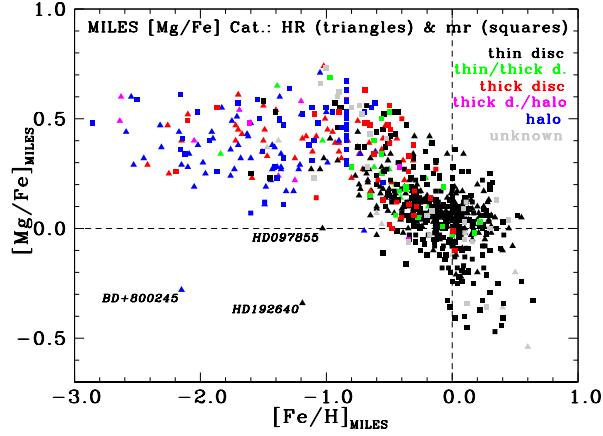}}
\caption{[Mg/Fe]$_{\rm MILES}$ vs. [Fe/H]$_{\rm MILES}$
(Galactic kinematic groups distinguished using distinct colours).}
\end{figure}

% Figure 4
\begin{figure}
\centerline{\includegraphics[width=5.90cm, angle=-90]{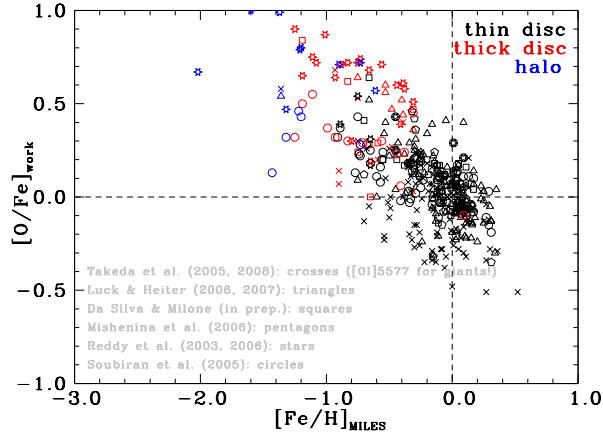}}
\caption{[O/Fe]$_{\rm work}$ vs. [Fe/H]$_{\rm MILES}$
(Galactic kinematic groups distinguished with different colours).}
\end{figure}

% Section 6 ------------------------------------------------------------------------------%
\section{Expansion of MILES}\label{s:maths}

We have compiled from {\sl HR} works a sample of candidate stars
to be incorporated into the MILES library.
The whole set contains around one thousand stars spread over both sky hemispheres
by adopting as reference the PASTEL stellar parameter catalogue \citep{2010A&A...515A.111S}.
The motivation is to improve the sample distribution over the MILES 4-D parameter space
({\it T}$_{\rm eff}$, log~$g$, [Fe/H], [Mg/Fe]).
Having excluded spectroscopic binaries, anomalous variable and peculiar stars,
we have defined a sample of around 400 field stars
that are observable with the Isaac Newton Telescope (INT)
at Observatorio del Roque de Los Muchachos (Canary Islands, Spain).
We observed 218 additional stars in two runs in 2011 adopting the same spectroscopic instrumental setup
employed for the original MILES library (INT with the Intermediate Dispersion Spectrograph).
The photospheric parameters of the additional sample have been calibrated onto the MILES system,
while [Mg/Fe] will be calibrated soon.

% Section 7 ------------------------------------------------------------------------------%
\section{Planning of SSP modelling}\label{s:maths}

The \citet{2010MNRAS.404.1639V} models show a Mg/Fe scaled-solar abundance ratio at solar metallicity
but increase it with decreasing metallicity following the abundance pattern of solar neighbourhood.
To improve the accuracy of these SSPs we intend to construct $\alpha$-enhanced models
with variable [$\alpha$/Fe] for a range in ages and metallicities.
We will build state-of-the-art SSP models with distinct [$\alpha$/Fe]
by selecting MILES stars according to their [Mg/Fe] values within a 4-D interpolation scheme,
whose spectra will be integrated along isochrones of similar [$\alpha$/Fe] under a self-consistent approach.
Figure 5 exemplifies how the matching of MILES stars with an isochrone
could be improved by also including the new data set and kinematically filtering the stars.

% Figure 5
\begin{figure}
\centerline{\includegraphics[width=5.637cm, angle=-90]{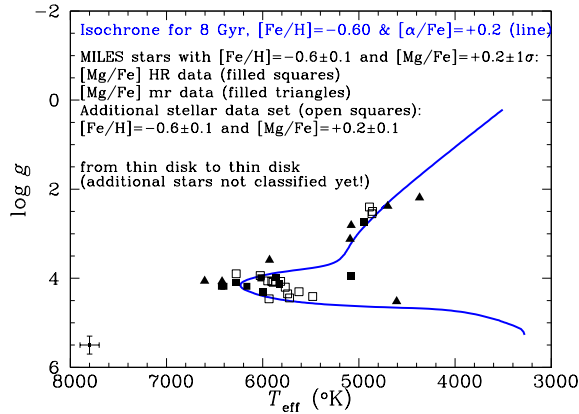}}
\caption{MILES stars for [Fe/H] = -0.6 and [Mg/Fe] = +0.2 from thin disk group only
cross-matched with a 8 Gyr isochrone of \citet{2008ApJS..178...89D}.}
\end{figure}

%------------------------------------------------------------------------------%
\section{Summary}\label{s:maths}

We have determined [Mg/Fe] with a precision of $\sim$0.1 dex for about 80\% of MILES stars
that are placed on a uniform scale (available on request or in Milone et al. 2011).
A robust spectroscopic analysis was carried out using the MILES spectra and LTE spectral synthesis of two Mg features.
This same semi-automated approach can be applied to recover [Ca/Fe] and perhaps [C/Fe].
MILES is currently being expanded by around 20\% through the inclusion of stars with known
{\it T}$_{\rm eff}$, log~$g$, [Fe/H] and [Mg/Fe]
that fills in some gaps of the library 4-D parameter space and increase the star density in other regions.
The transformation of their photospheric parameters to the MILES homogeneous system
has been carried out, and the calibration of their [Mg/Fe] will be completed soon.
The abundances of calcium, carbon, nitrogen and oxygen are currently being investigated for the MILES stars
(compilation from literature for all of them and Ca atomic features spectroscopically analysed at mid-resolution).
Specifically carbon and nitrogen most likely play an important role in the blue spectral region
as we have concluded from our empirical analysis based on MILES spectral ratios
between pairs of similar stars in {\it T}$_{\rm eff}$ and log~$g$. 
The Galactic kinematic classification of MILES stars with compiled [Mg/Fe] has been just computed
such that this information can be included in the SSP modelling.
Careful comparisons of the theoretical stellar predictions of the Lick indices against the MILES data
have revealed the good behaviour of Fe-sensitive indices predictions,
while highlighting areas for improvement in some models for the higher order H-Balmer features.
We intend to compute a set of self-consistent semi-empirical SSP models
with variable $\alpha$-enhancement for a range in ages and metallicities around solar.

%------------------------------------------------------------------------------%
%\section{References -- using ADS}\label{s:ADS}

%------------------------------------------------------------------------------%
\section*{Acknowledgements}

Milone thanks the Brazilian foundation FAPESP (2013/18393-9)
and the institutional programme PCI/MCTI/INPE (BSP 17.0018/2010-5, BSP 17.0061/2012-4).
Milone and Sansom thank FAPESP (2012/04953-0).
We thank R. Cacho, I. Mart\'{\i}n and E. M\'armol-Queralt\'o who helped us in one of the INT runs
as well as B. C. da Silva who is dealing with Ca feature spectral synthesis (PIBIC-INPE/MCTI 129330/2013-2).

%------------------------------------------------------------------------------%
% bibliography: produced from ADS using custom format of                       %
%                                                                              %
%     %z132 \\bibitem[%\2%(y)%\3m]%{R}\n   %\8.1g,%\Y,%\q,%\V,%\ p             %
%------------------------------------------------------------------------------%

\end{document}